\documentclass[sigconf]{acmart}%

\pdfoutput=1

\AtBeginDocument{%
  \providecommand\BibTeX{{%
    \normalfont B\kern-0.5em{\scshape i\kern-0.25em b}\kern-0.8em\TeX}}}

\copyrightyear{2024} 
\acmYear{2024} 
\setcopyright{rightsretained} 
\acmConference[CHI EA '24]{Extended Abstracts of the CHI Conference on Human Factors in Computing Systems}{May 11--16, 2024}{Honolulu, HI, USA}
\acmBooktitle{Extended Abstracts of the CHI Conference on Human Factors in Computing Systems (CHI EA '24), May 11--16, 2024, Honolulu, HI, USA}
\acmDOI{10.1145/3613905.3651047}
\acmISBN{979-8-4007-0331-7/24/05}

\usepackage[inline]{enumitem}
\usepackage{tikz}
\newcommand{\cir}[1]{\tikz[baseline]{%
\node[anchor=base, draw, circle, inner sep=0, minimum width=1.0em, thick]{\small #1};}}

\usepackage{graphicx}
\graphicspath{{./fig/}}

\begin{document}

\title[%
What Does Evaluation of Explainable Artificial Intelligence Actually Tell Us?%
]{%
What Does Evaluation of Explainable Artificial Intelligence Actually Tell Us?%
}

\subtitle{A Case for Compositional and Contextual Validation of XAI Building Blocks}

\author{Kacper Sokol}
\email{kacper.sokol@inf.ethz.ch}
\orcid{0000-0002-9869-5896}
\affiliation{%
  \institution{Department of Computer Science, ETH Zurich} %
  \country{Switzerland}
}

\author{Julia E.\ Vogt}
\email{julia.vogt@inf.ethz.ch}
\orcid{0000-0002-6004-7770}
\affiliation{%
  \institution{Department of Computer Science, ETH Zurich} %
  \country{Switzerland}
}

\begin{abstract}
Despite significant progress, evaluation of explainable artificial intelligence remains elusive and challenging. %
In this paper we propose a fine-grained validation framework that is not overly reliant on any one facet of these sociotechnical systems, and that recognises their inherent modular structure: technical building blocks, user-facing explanatory artefacts and social communication protocols. %
While we concur that user studies are invaluable in assessing the quality and effectiveness of explanation presentation and delivery strategies from the explainees' perspective in a particular deployment context, the underlying explanation generation mechanisms require a separate, predominantly algorithmic validation strategy that accounts for the technical and human-centred desiderata of their (numerical) outputs. %
Such a comprehensive sociotechnical utility-based evaluation framework could allow to systematically reason about the properties and downstream influence of different building blocks from which explainable artificial intelligence systems are composed -- accounting for a diverse range of their engineering and social aspects -- in view of the anticipated use case. %
\end{abstract}

\ccsdesc[500]{Human-centered computing~User studies}
\ccsdesc[500]{General and reference~Evaluation}
\ccsdesc[500]{General and reference~Validation}
\ccsdesc[300]{General and reference~Experimentation}
\ccsdesc[300]{General and reference~Metrics}
\ccsdesc[100]{Computing methodologies~Machine learning}
\ccsdesc[100]{Computing methodologies~Artificial intelligence}

\keywords{%
Evaluation; %
Validation; %
User Studies; %
Explainability; %
Interpretability; %
Machine Learning; %
Artificial Intelligence. %
}%

\maketitle

\section{Evaluation Purposes}%

Predictive models based on Artificial Intelligence (AI) algorithms come with a myriad of well-defined (predictive) performance metrics that guide us through their development, validation and deployment. %
When it comes to assessing their interpretability or explainability, however, we lack agreed-upon evaluation criteria, protocols and frameworks. %
While the body of literature addressing this fundamental challenge is steadily growing, this aspect of predictive algorithms is elusive and notoriously difficult to define and measure, which can be attributed to its sociotechnical situatedness~\cite{rudin2019stop,sokol2021explainability,ehsan2023charting,keenan2023mind,sokol2023reasonable}. %
Since eXplainable Artificial Intelligence (XAI) approaches are inextricably intertwined with human agents -- for whom the output of such methods is often intended~\cite{miller2019explanation} -- relevant evaluation procedures need to account for purely \emph{technical} as well as \emph{social properties}~\cite{sokol2020explainability,doshi2017towards}. %
This duality reflects the recent transition of XAI from a predominantly engineering field towards a more human-centred endeavour that accounts for explainees both when building and assessing XAI techniques~\cite{miller2019explanation}. %
With this shift, a new evaluation and validation paradigm has emerged, in which \emph{user studies} -- currently considered the gold standard -- organically became the go-to vehicle to assess various characteristics of explainability methods, complementing the pre-existing algorithmic metrics. %

User-based validation is indispensable for judging informativeness, acceptability, usability and the like of algorithmic explanations, however the rapid adoption of this evaluation strategy without appropriate foundational work and clear guidelines has resulted in a chaotic, confusing and difficult to navigate findings that do not necessarily exhibit \emph{ecological validity} and cannot be easily compared. %
For example, researchers tend to reuse (questionable) user study protocols found in prior work, or propose new, ad-hoc evaluation setups. %
Surveys published in this space are increasingly common and attempt to abate such issues by systematising the landscape of human- and model-centred evaluation approaches, but they fall short of proposing a flexible framework that can be adapted to unique explainability scenarios in a principled manner~\cite{zhou2021evaluating,mohseni2021multidisciplinary,hoffman2018metrics,herman2017promise}. %
This is not to say that such guidelines can be easily designed or are even possible for a generic explainer and application domain; %
nonetheless, lack of foundational research on technical and social aspects of XAI evaluation as well as various interdependencies between the two impede further progress. %
XAI is a complex construct that strives to bridge the \emph{sociotechnical gap}~\cite{ackerman2003sociotechnicalgap,ehsan2023charting}, hence it is unlikely to have a single prescriptive evaluation formula. %

In view of these shortcomings, we review how interpretability and explainability are evaluated, looking into algorithmic and user-centred approaches as well as frameworks built on top of them (Section~\ref{sec:approaches}). %
This analysis leads us to five observations (Section~\ref{sec:pillars}). %
\begin{enumerate*}[label=\protect\cir{\arabic*}]
\item Evaluation of XAI approaches sometimes yields inconsistent or contradictory results, which is especially prevalent for user-based validation. %
While this is problematic from the practical point of view, it is unsurprising given the sociotechnical situatedness of such systems and (slight) variations across the employed evaluation protocols. %
\item Context in which explainability takes place is especially important in this respect and it is likely to affect otherwise identical user-based evaluation studies; nonetheless, it is often neglected or ignored. %
\item The complexity of the human explanatory process -- extensively covered by psychology and cognitive sciences research -- further exacerbates this issue. %
While explainability is a social, iterative and interactive process, such a perspective tends to be overlooked in the evaluation literature. %
\item Additionally, despite being fundamentally modular, XAI systems are often treated as monolithic tools -- including for evaluation purposes -- arguably leading to suboptimal and misaligned explanatory setups. %
\item In addition to considering the trustworthiness of XAI tools, one should also account for the quality of the model being explained since it can skew evaluation results, which is rarely acknowledged. %
\end{enumerate*}

In view of these observations, shifting towards human-centred evaluation of XAI approaches should go beyond simply computing human-compatible metrics or executing user studies, and instead strive to better understand the sociotechnical gap that explainers try to bridge~\citep{keenan2023mind}. %
Achieving this goal is likely to require a fundamental change in how we approach evaluation of explainable artificial intelligence. %
To lay the foundation for this transition we propose a sociotechnical utility-based XAI evaluation framework (Section~\ref{sec:discussion}) whose three tenets are grounded in our extensive literature analysis. %
\begin{enumerate*}[label=\protect\cir{\arabic*}]
    \item We should recognise functionally independent (and interoperable) components of explainability systems -- such as core building blocks of algorithmic explainers that produce explanatory insights as well as artefacts, interfaces and communication protocols used to present them and deliver them to users -- some of which elements are predominantly technical while other social. %
    \item Given this compositionality of XAI approaches, explainers should not only be custom-built for the problem at hand but also evaluated on multiple levels -- in terms of their individual components, groups thereof and holistically as end-to-end systems -- choosing appropriate, algorithmic, social or mixed, validation mechanisms in each case. %
    \item The envisaged operational context should thus determine the most appropriate evaluation criteria, metrics and protocols since certain findings will only be valid within the chosen evaluation setting, while others will generalise beyond it. %
\end{enumerate*}

Operating within this evaluation framework could give rise to a comprehensive library of standalone XAI building blocks that are thoroughly tested and validated in well-defined contexts. %
Such a repository can then be used to compose bespoke explainers that are trustworthy and appropriate for the task at hand, which systems themselves can be tested end-to-end and contributed back to the library. %
The framework also promises to streamline evaluation. %
For example, ante-hoc interpretability approaches would only require context-specific social validation since their outputs are inherently truthful~\cite{rudin2019stop}; %
counterfactual explainability algorithms would only need to be assessed based on the correctness of their output and domain-specific (quantitative) desiderata since this explanation type has already been thoroughly tested with humans~\cite{miller2019explanation}; and %
explanation presentation media and communication protocols would only be tested on their own in well-defined contexts given their functional and operational independence from (algorithmic) explanation generation mechanisms. %
Following the discussion of our framework, we finish the paper by summarising our contribution, drawing conclusion and exploring future work (Section~\ref{sec:conclusion}). %

\section{Evaluation Approaches\label{sec:approaches}}%

An ever increasing abundance of XAI methods brought a need for their systematic assessment and principled comparison~\citep{sokol2020explainability,markus2021role}, which up to a point had been limited to ``I know it when I see it'' evaluation approaches~\citep{doshi2017towards}. %
To this end, researchers have devised numerous algorithmic and (subjective or objective) user-based evaluation techniques as well as frameworks built on top of them, whose overviews, taxonomies and surveys are widely available~\citep{narayanan2018humans,hoffman2018metrics,sokol2020explainability,mohseni2021multidisciplinary,markus2021role,zhou2021evaluating,coroama2022evaluation,nauta2022anecdotal,byrne2023good}. %
Nonetheless, despite significant progress, better evaluation methodologies, frameworks and workflows are still very much needed~\citep{longo2023explainable}. %

\paragraph{Evaluation Frameworks}
XAI taxonomies and lists of desiderata spanning social and technical dimensions -- such as \emph{Explainability Fact Sheets}~\cite{sokol2020explainability} -- strive to elicit properties required of explanations~\citep{narayanan2018humans}, however they stop short of providing specific evaluation metrics. %
More commonly, XAI techniques are evaluated directly with %
\begin{itemize}[font=\normalfont\itshape]%
    \item \emph{heuristic-based}, %
    also called objective, computational and algorithmic, and %
    \item \emph{user-based}, %
    also called human-centred, %
\end{itemize}%
 metrics that capture subjective and objective quantitative and qualitative properties of explanations~\citep{bibal2016interpretability,vilone2020explainable}. %
A complementary viewpoint categorises evaluation approaches into %
\begin{itemize}[font=\normalfont\itshape]%
    \item \emph{functionally-grounded} %
    -- no humans and (algorithmic) proxy tasks, %
    \item \emph{human-grounded} %
    -- real humans and simplified tasks, and %
    \item \emph{application-grounded} %
    -- real humans and real tasks, %
\end{itemize}%
types, with the first one loosely corresponding to heuristic-based and the latter two linked to user-based~\citep{doshi2017towards}. %

Quantitative metrics, pertaining to both algorithmic and social aspects of explainability systems, are often categorised according to the element of the XAI process that they target~\citep{nauta2022anecdotal}: %
\begin{itemize}[font=\normalfont\itshape]%
    \item \emph{explanation content}, %
    e.g., correctness, completeness, consistency, continuity, contrastivity and covariate complexity; %
    \item \emph{explanation presentation}, %
    e.g., compactness, compositionality and confidence; and %
    \item \emph{explanation recipient}, %
    e.g., context, coherence and controllability. %
\end{itemize}%
A more comprehensive framework delineates evaluation approaches according to the aspect of the XAI process that they target: %
\begin{itemize}[font=\normalfont\itshape]%
    \item \emph{explanatory information} %
    output by an explainer,
    \item \emph{user understanding}, %
    and
    \item \emph{explainability desiderata}, %
\end{itemize}%
thus additionally accounting for the taxonomy perspective~\citep{speith2023new}. %

Another viewpoint separates %
\begin{itemize}[font=\normalfont\itshape]%
    \item \emph{computational interpretability}, i.e., \emph{numerical} methods for evaluating XAI algorithms, from %
    \item \emph{human interpretability}, i.e., \emph{human-subjects evaluation} for assessing all the other aspects of the explanatory pipeline, %
\end{itemize}%
leading to a distinction between \emph{interpretable artificial intelligence} and \emph{explainable interfaces}~\citep{mohseni2021multidisciplinary}; %
these two concepts correspond to explanatory insights extracted from a data-driven model and their presentation to an explainee respectively. %
Simply put, this split enables measuring the intrinsic goodness of an explanation independently from the explainee's satisfaction~\cite{holzinger2020measuring}. %
These two components can be assessed through different \emph{types of evaluation} -- computational, qualitative and quantitative with human-subjects -- that span diverse \emph{evaluation measures}, e.g., user trust, satisfaction, mental model and understanding as well as explanation usefulness, task performance and strictly computational measures that include correctness, completeness and fidelity. %
Such a comprehensive evaluation toolbox allows to explicitly account for the audience and application domain, whose specific explanatory needs and goals should inform the choice of an XAI approach to ensure its effectiveness, in which case any evaluation results may not generalise beyond the specific testing setup and context. %

Evaluation approaches can alternatively be categorised according to~\citep{donoso2023towards}: %
\begin{itemize}[font=\normalfont\itshape]%
    \item the \emph{property} they measure; %
    \item the \emph{element of an explanation} they target; and %
    \item the evaluation \emph{procedure type}, e.g., questionnaires, metrics and the like. %
\end{itemize}%
Within this framework each measure applies to only one explanation \emph{element} -- its generation, abstraction, format or communication -- but a single \emph{property} can be assessed with multiple measures (possibly based on different \emph{procedures}); %
this decoupling facilitates choosing measures appropriate for the explainability context and desiderata. %
The Predictive / Descriptive / Relevant (PDR) explainable AI evaluation framework, on the other hand, distinguishes metrics that capture~\cite{murdoch2019definitions}: %
\begin{itemize}[font=\normalfont\itshape]%
    \item \emph{predictive accuracy}, i.e., the performance of the explained model; %
    \item \emph{descriptive accuracy}, i.e., the degree to which explanations are representative of and truthful with respect to that model; and %
    \item \emph{relevance}, i.e., the appropriateness of explanatory insights in the context of the anticipated audience and application domain. %
\end{itemize}%
The distinction between various objectives or purposes of an explanation with respect to an AI model %
can also guide XAI evaluation~\citep{yao2021explanatory}; %
the possible goals are: %
\begin{itemize}[font=\normalfont\itshape]%
    \item \emph{diagnostic} -- to expose inner workings, %
    \item \emph{explication} -- to understand locally, %
    \item \emph{expectation} -- to understand globally, and %
    \item \emph{functional} (role-driven) -- to contextualise the operation and justify the use. %
\end{itemize}%
Overall, XAI evaluation approaches and procedures that consider \emph{predictive performance} of the explained model, capture \emph{social desiderata}, are \emph{context-aware}, or account for the (multi-step) \emph{interaction} between users and explainers are largely missing~\citep{murdoch2019definitions,donoso2023towards,speith2023new} despite the social aspects of explainability being a key tenet of human-centred XAI~\citep{miller2019explanation}. %

\paragraph{Algorithmic Evaluation}%
Computational measures of explainability predate the current surge of interest in XAI. %
For example, explanation \emph{fidelity}, \emph{accuracy}, \emph{consistency} and \emph{comprehensibility} have already been used in \citeyear{andrews1995survey} to evaluate rule-based models~\citep{andrews1995survey}. %
Since then many more such metrics have been proposed~\citep{nauta2022anecdotal} and implemented in comprehensive benchmarking suites~\cite{agarwal2022openxai}. %
While predominantly these evaluation approaches measure \emph{technical} properties of \emph{individual} explanations, other techniques exist that deal with groups of explanations (e.g., their diversity), focus on data domain- or user-specific desiderata (e.g., sparsity, feasibility and actionability), or target particular types of explanations and explainer families (e.g., counterfactuals and post-hoc methods)~\citep{zhou2021evaluating}. %

An alternative categorisation of this type of XAI evaluation metrics is based on their operationalisation. %
Some measures are premised on the concept of \emph{robustness}. %
Here, two common themes are %
\begin{enumerate*}[label=\protect\cir{\arabic*}]
    \item
to remove features that are important according to an explanation, or inject noise or perturbation to these features, and measure the effects of such a procedure on the performance of the explained model; and %
    \item
to study the stability and invariance of explanations with respect to different implementations of a predictive model or randomisations of its parameters~\citep{hooker2018evaluating,murdoch2019definitions,vilone2020explainable,lee2022heatmap}. %
\end{enumerate*}

Other types of measures \emph{establish a reference point}. %
Here, some evaluation approaches investigate whether similar data points yield comparable explanations for a fixed predictive model and explainer; or whether a single instance is explained with comparable insights produced by multiple explainers for a fixed predictive model; as well as other variations of such a setup~\citep{alvarez2018robustness}. %
Approaches based on \emph{ground truth} -- given by synthetic data or annotation of real data sets -- constitute a third category. %
Here, we can build a transparent model and compare its internals to explanations produced by an XAI technique; %
evaluate explanatory insights against human annotations of a data set, e.g., image regions marked as important for a given class; or %
assess explanations with respect to a known data generation process used to produce synthetic data on which an AI model is trained~\citep{murdoch2019definitions,guidotti2021evaluating,hoffman2018metrics,yang2019benchmarking,mohseni2021quantitative,zhou2022feature}. %

\paragraph{User-centred Evaluation}%
From an explainee's perspective, the quality and effectiveness of an explanation can be measured along the dimensions of satisfaction, mental model state, task performance, trust and correctability (i.e., the ability to identify and rectify errors of a predictive system)~\citep{gunning2017explainable}. %
More broadly, user-based evaluation aims to assess the pertinence of the explainee's degree of understanding with respect to a specific task after being exposed to an explanation~\citep{bau2017network,kim2021multi,sokol2021explainability}. %
This property can be measured \emph{subjectively} -- e.g., user-reported satisfaction, trust, confidence, preference and understanding -- and \emph{objectively} -- e.g., based on problem solving and task performance (appraised via correctness and completion time) or physiological and behavioural factors~\citep{zhou2021evaluating}. %
The two assessment strategies can nonetheless yield contradictory results~\citep{herm2021don}. %

One factor contributing to this observed disagreement %
as well as diminishing the value of \emph{subjective} evaluation is %
the susceptibility of this approach to %
producing volatile and misleading results %
since an explanation can make a user \emph{falsely} believe -- through persuasion or otherwise -- that they understand the underlying predictive system~\citep{byrne2023good}. %
This phenomenon has been famously demonstrated by the \emph{placebic} explanations used in the \emph{copy machine} study~\citep{langer1978mindlessness}. %
In contrast, %
objective measures of explainees' factual or procedural knowledge and understanding can quantify the \emph{retention}, \emph{transfer} and \emph{recall} of explanatory information~\citep{herm2021don} by, for example, quizzing them about model errors, predictions and decision boundaries~\cite{rosenfeld2019explainability,chen2022machine,spreitzer2022evaluating,morrison2023eye,xuan2023can}. %
Many more objective and subjective user-centred XAI evaluation criteria can be found in the literature~\citep{hoffman2018metrics,lage2019evaluation,mohseni2021multidisciplinary,hoffman2023measures}. %

\section{Evaluation Deficiencies\label{sec:pillars}}%

Having reviewed the state of the art in XAI evaluation, we shift our attention to the applied aspects of this process. %
The literature survey has led us to five important observations that go beyond the general disagreement on what metrics to use and how to set up evaluation. %
Specifically, these are \emph{inconsistent evaluation findings}, the role of \emph{operational context} in explainability, the significance of the \emph{human explanatory process}, the challenges of the \emph{inherent complexity of algorithmic explainers} and the implications of the varied \emph{quality of the explained model}. %

\paragraph{Inconsistent Evaluation Findings}%
While explainability strives to provide important insights into the functioning of AI models to the benefit of their users, this objective may not always be attained. %
XAI evaluation literature reports an array of inconsistent, counterintuitive and contradictory findings both within the scope of an individual study and across different studies. %
In particular, explainability has been shown to offer less than desired or anticipated result, have no effect at all, or even be detrimental, e.g., hurt task completion~\cite{jesus2021can,suresh2022intuitively}. %
The literature documents common cases where explainees -- lay people as well as AI experts -- misuse or misinterpret insights output by sate-of-the-art explainers, leading to \emph{negative comprehension} exhibited by wrong choices, incorrect beliefs or excessive trust and reliance on predictive systems~\cite{kaur2020interpreting,small2023helpful,xuan2023can}. %
For example, the (ante-hoc) interpretability of inherently transparent models may be problematic or ineffective, which was shown to be the case for linear models and decision trees~\cite{bell2022its,sokol2023reasonable}; %
the formulation of the former was found to be confusing, and the visualisation of the latter to mislead explainees into selecting the root node feature as the most important one. %
Furthermore, realistic, application-grounded XAI evaluation attempts -- which recruit real humans, e.g., domain experts, and rely on real models trained on real data to solve real tasks -- can yield disparately different results from those offered by simplified setups~\cite{amarasinghe2022importance}. %

\paragraph{Neglected Explainability Context}%
With the shift towards human-centred XAI, the stakeholders -- including their requirements, needs, goals, knowledge and expertise -- have become an important consideration when designing explainability systems~\cite{miller2019explanation,tomsett2018interpretable,preece2018stakeholders,mueller2021principles}. %
As some researchers note, an explanation can be correct but at the same time it may not be any good or useful (for a given task)~\cite{achinstein1983nature}, or simply put ``one explanation does not fit all''~\cite{sokol2020one}. %
To be effective, explanations should align with the explainees' objectives and provide them with the required information, thus be helpful in a given situation as outside of it the explanatory insights may simply become irrelevant or meaningless~\cite{mueller2021principles,cabour2022explanation}. %
This crucial role of context, nonetheless, seems to be rarely acknowledged when choosing metrics and protocols for evaluating XAI techniques, which appears to be one of the main reasons for the inconsistent findings highlighted in the previous paragraph. %
Notably, distinct evaluation criteria may be assigned different importance depending on the XAI application area, thus guide the design of bespoke explainability methods as well as their custom evaluation strategies~\cite{liao2022connecting}. %

These observations lead to two conclusions: %
\begin{enumerate*}[label=\protect\cir{\arabic*}]
    \item XAI techniques should be designed to address real-world use cases that ground them in well-defined practical requirements; and %
    \item their evaluation process should be adapted to and executed in a realistic setting that reflects the anticipated deployment scenario, i.e., it should be \emph{application-grounded}, since such results tend not to generalise beyond the specific evaluation context~\cite{jesus2021can,amarasinghe2022importance,mueller2021principles}. %
\end{enumerate*}
Recognising these principles is especially important for user-based evaluation since it has high execution costs and cannot be easily repeated under a revised experimental setup, unlike algorithmic evaluation where a number of additional metrics can be computed post-hoc without much overhead. %
While a recent initiative to streamline \emph{user-based} evaluation by replacing real humans with \emph{large language models} strives to overcome this challenge and provide a means to test XAI systems with ``different audiences''~\citep{chen2021simulated}, this approach is unlikely to produce meaningful results given the richness and uniqueness of each XAI deployment context. %

\paragraph{Oversimplified Human Explanatory Process}%
Explainability is not a \emph{property} of an explanatory artefact per se, but rather an (interactive and iterative communication) \emph{process} by which an explainer and explainee establish common ground~\cite{achinstein1983nature,miller2019explanation,mueller2021principles,sokol2021explainability}. %
Triggered by expectation failures (e.g., a surprising event), learning needs or task requirements, it allows to adjust misconceptions, reconcile differences and resolve disagreements by helping a person to develop correct knowledge, beliefs and understanding of the explained phenomenon~\cite{gregor1999explanations}. %
Such a perspective entails a separation between %
\begin{itemize}[font=\normalfont\itshape]%
    \item
the \emph{explanatory artefacts}, i.e., the type, content, presentation format and provision mechanism of an explanation determined by \emph{what needs to be explained}, and %
    \item
the \emph{act of explaining}, i.e., the explanatory process governed by \emph{how to explain it}, %
\end{itemize}%
which draws an implicit distinction between user interfaces and explainability~\cite{achinstein1983nature,langer2021we,kaur2022sensible,cabour2022explanation,keenan2023mind,gregor1999explanations}. %

\begin{figure*}[t]
    \centering
    \includegraphics[width=.80\linewidth]{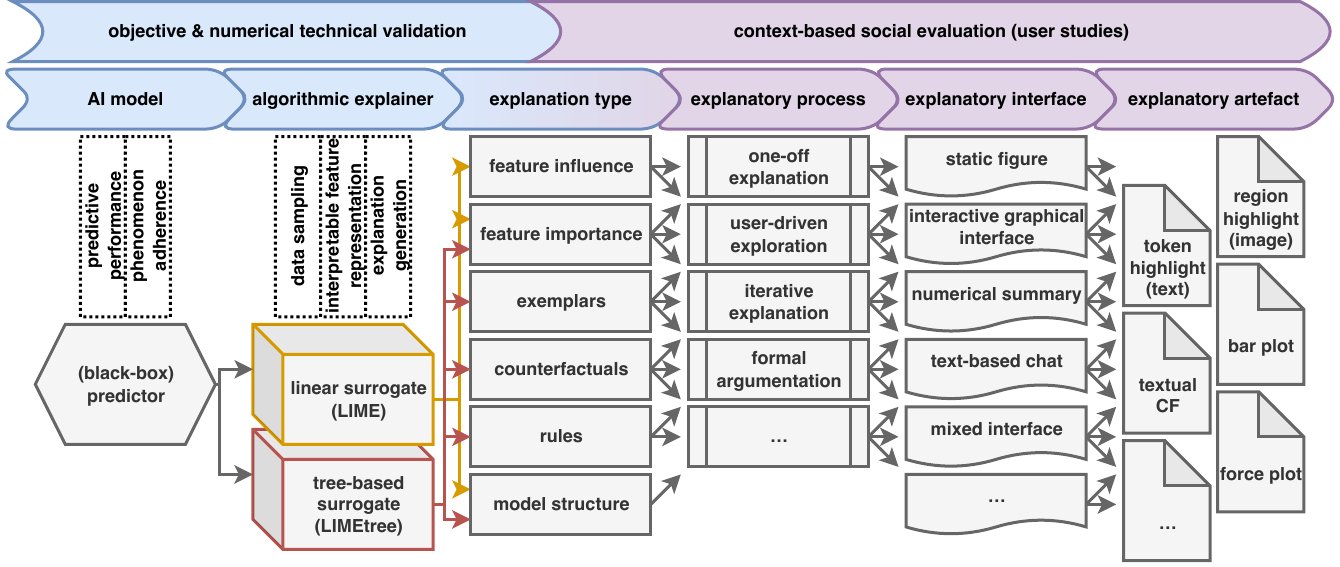}%
    \caption{%
    Depiction of our (rudimentary) sociotechnical utility-based evaluation framework illustrating functionally independent building blocks of XAI systems. This example uses linear (LIME~\citep{ribeiro2016why}) and tree-based (LIMEtree~\citep{sokol2020limetree}) surrogate explainers. %
\label{fig:example}}
\end{figure*}

This categorisation is particularly useful when trying to identify the best XAI system for the task at hand. %
For example, in certain cases explanations that are \emph{true enough} may suffice and serve their purpose well; %
such as a simplified depiction of an underground rail network that portrays stops and transfer points without spatial scale and orientation~\cite{paez2019pragmatic}. %
The presentation and communication of an explanation also play a vital role since different representations or framings can alter -- expedite or impede -- human perception, reasoning, problem-solving and decision-making, both in general and in the context of a particular task~\cite{zhang1994representations,herm2021don,huysmans2011empirical,kuhl2023better}. %
For example, Arabic numerals are more efficient for arithmetic than Roman numerals, and digital clocks are better suited for determining precise time while analog clocks facilitate easier tracking of time intervals (as one does not need to perform calculations), despite the two representations being isomorphic in both cases~\cite{larkin1987diagram,norman1993cognition,patel2013cognitive}. %
In terms of evaluation, given the importance of explanation presentation some researchers have recently begun to explore how different interfaces influence the perception of a single explainability algorithm~\cite{mucha2021interfaces}. %
While a wide selection of evaluation approaches exists for \emph{explanatory artefacts} (reviewed earlier in Section~\ref{sec:approaches}), methods that target the \emph{act of explaining} -- which can possibly be an interactive and iterative, multi-turn process -- are still largely missing. %

\paragraph{Misconstrued Explainer Structure}%
While the separation between explanation content, presentation format, provision mechanism and the act of explaining is gaining traction in XAI, explainability systems remain predominantly perceived as monolithic entities despite in fact being highly modular~\cite{sokol2022what}. %
This misconception results in a widespread use of off-the-shelf explainers without much thought or effort put into customising these tools and building bespoke XAI systems on top of them, which practice promises to offer more trustworthy and better quality explanations~\cite{suresh2022intuitively}. %
For example, in addition to different explanation presentation formats, LIME~\cite{ribeiro2016why}, a popular surrogate explainer, has three functionally independent algorithmic components -- interpretable feature representation, data sampling and explanation generation -- each of which has a number of parameters itself~\cite{sokol2019blimey}. %

While many studies assess the effectiveness of an explanation type, content and presentation medium, they often overlook the inherent correctness and trustworthiness of the underlying explanatory insight, or rather the pervasive lack thereof~\cite{yang2019xfake,linder2021how,rudin2019stop}. %
This inspires an alternative, \emph{diagnostic} conceptualisation of XAI, which focuses on providing users with rigorously tested and well-specified insights into a predictive model instead of attempting to solve the ill-defined ``black box'' problem~\cite{chen2022interpretable}. %
In terms of evaluation, we should thus not only test explainers end-to-end but also validate their individual components independently and provide clear guidelines on how and when to operationalise these tools to guarantee correctness and trustworthiness of their outputs~\cite{nguyen2020quantitative,sokol2022what}. %

\paragraph{Overlooked Model Quality}%
Another aspect of XAI evaluation that is rarely considered in the literature is the interplay between the quality of a data-driven model -- e.g., as determined by its truthfulness with respect to the underlying data-generating phenomenon or, simply, predictive performance -- and the effectiveness of an explainer, which relation may not be fixed across AI systems with different capabilities~\cite{papenmeier2022complicated}. %
Such considerations give rise to a distinction between \emph{scientific} and \emph{algorithmic explanations} determined by the scientific consistency of the explained predictive model, the lack of which may result in untruthful explanations that can be used to manipulate or persuade an explainee, or justify a preselected conclusion~\cite{baumeister1994self,roscher2020explainable}. %
\emph{Algorithmic} explanations can be linked to \emph{explanatory debugging} or (the aforementioned) \emph{model diagnostics} whereby XAI methods serve as inspection tools for peering inside, identifying shortcomings and improving predictive systems~\citep{kulesza2015principles,kim2021multi,yao2021explanatory}. %

\section{Sociotechnical Utility-based Evaluation Framework\label{sec:discussion}}%

To address the XAI evaluation challenges identified in the previous section we propose a \emph{sociotechnical utility-based evaluation framework} that recognises operationally independent components of explainability systems and maps out their relations. %
Specifically, it separates \emph{predictive models} and the \emph{algorithms that extract explanatory insights} -- i.e., the technical constructs -- from \emph{explanatory artefacts} and \emph{interfaces} that present these insights to explainees -- i.e., the social elements -- as well as \emph{communication protocols} that bridge the two sides -- i.e., the sociotechnical layer. %
This division can be compared to %
the distinction between \emph{data analytics} and \emph{information visualisation}, which naturally delineates the process of extracting insights from data and presenting them to diverse audiences~\cite{mccandless2012information}. %
Within our framework the questions of \emph{what information to present} and \emph{how to deliver it} become virtually independent of the challenges of \emph{how to extract this information} and \emph{its objective validity}. %
This separation arises naturally from the \emph{utility} perspective where we are primarily concerned with the \emph{functional role} of each XAI building block, which in turn determines the objectives of its evaluation. %
Figure~\ref{fig:example} offers a rudimentary example of our framework for surrogate explainers. %

\begin{figure}[t]
    \centering
    \includegraphics[width=.520\linewidth]{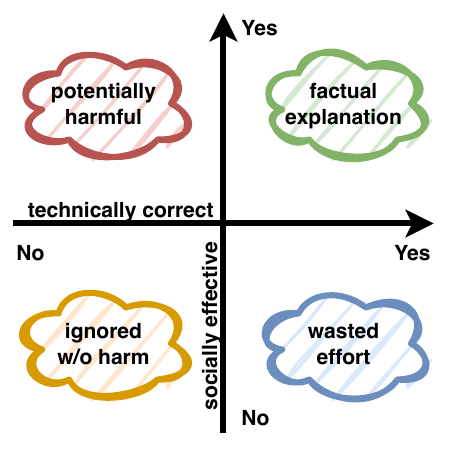}%
    \caption{%
High-level view of our utility-based evaluation framework that separates %
\emph{technical} correctness of explanatory insights from \emph{social} effectiveness of their delivery. %
\label{fig:evaluation-grid}}
\end{figure}

This decoupling allows to account for any deficiencies of the predictive model and its faithfulness to the underlying data-generating phenomenon, thus deal with cases where explanations are produced for truthful and correct predictions as well as outputs with lacklustre confidence and those that are outright wrong. %
It also takes into consideration the validity of explanatory insights with respect to the predictive model, which enables differentiating between \emph{descriptive} -- i.e., truthful, factual and correct -- and \emph{persuasive} -- i.e., familiar, easy to understand and convincing -- explanations~\cite{herman2017promise}. %
This distinction is important as the act of providing an explanation can itself engender trust in a predictive model even if the underlying explanatory artefact conveys no meaningful or correct information~\cite{langer1978mindlessness}. %
An explanation can therefore be \emph{dangerous} and \emph{harmful} if it is effective but incorrect, or a \emph{wasted effort} when the opposite is true -- this high-level relation is captured in Figure~\ref{fig:evaluation-grid}. %
Notably, an explanation can be fundamentally correct with respect to a predictive model yet considered flawed by the recipient because of an inconsistency between the data patterns used by the model and the explainee's perception of the underlying phenomenon. %
This scenario may arise either when the algorithmic decision is based on spurious correlations or when the user's mental model is flawed, which situations can be easily distinguished within the proposed framework. %

Generic algorithmic evaluation or user study protocols that treat explainers as end-to-end systems may not elicit detailed enough information to attribute the success or failure of such tools to any of their individual building blocks. %
For example, XAI methods tend to provide insights of a single \emph{type} that can often be presented with a diverse range of explanatory artefacts, %
but also different explainers can share an identical explanation type and explanatory artefact, possibly causing confusion among explainees. %
Additionally, explanations may be very effective, appealing and easy to understand but their content, i.e., the explanatory insight, may still be incorrect -- or vice versa -- as noted in the previous paragraph. %
The proposed framework allows to minimise such implicit interactions between individual components of XAI systems during their evaluation. %
To this end, the intended function of each building block as well as its role in distinct higher-level modules and the entire end-to-end explainer are examined to determine the most appropriate evaluation criteria, measures and approaches at each stage. %
Such a practice can be compared to testing in software engineering where each functionally independent component is assessed on its own in isolation (unit tests), as a part of bigger modules (integration tests), as a cog of an entire application (system tests) and on a meta-level in terms of said application fulfilling the users' needs (acceptance tests). %

Within our evaluation framework, proposing a new XAI approach for a well-established type of an explanation -- such as counterfactuals -- only requires validating its algorithmic properties, e.g., correctness with respect to the explained predictive model and selected user-centred technical properties, %
whereas introducing new types of an explanation or novel explanatory artefacts and communication mechanisms entails user-based evaluation. %
This is particularly desirable for ante-hoc explainers since they are guaranteed to produce faithful explanatory insights, which expedites their testing~\cite{rudin2019stop,sokol2023reasonable}. %
Such an approach empowers XAI researchers and practitioners to focus on evaluating the core aspect of each of their individual contributions, regardless of whether it is an explainability algorithm, artefact, communication protocol or end-to-end system comprising of all the three components. %
In particular, it absolves designers from running multiple superfluous user studies, prompts them to better specify the objective of such experiments and helps to prevent error-prone explainability algorithms from biasing study results. %
Our framework also highlights the urgent need for evaluation approaches that are capable of assessing the social and sociotechnical properties of explainers given that interactive and iterative explainability is the key tenet of human-centred XAI. %

The separation of XAI components offered by our framework additionally allows each evaluation task to account for the anticipated deployment context: application domain, target audience, explanation role as well as many other factors. %
Therefore, expensive to run user studies and focus groups can be reserved for assessing the effectiveness of a given explanation type conveyed by a specific artefact and communicated through a particular explanatory process in a well-defined \emph{operational context} since such results do not necessarily transfer across different use cases as noted earlier. %
Recognising independent XAI components and documenting the setup and results of their evaluation can also give rise to a library of thoroughly-tested and well-understood explainability building blocks as well as their end-to-end compositions. %
Such a comprehensive repository would facilitate building bespoke XAI systems from components that are appropriate for the task at hand, resulting in more reliable explainers and better user experience. %

\section{Conclusion and Future Work\label{sec:conclusion}}%

In this paper we introduced a utility-based evaluation framework of XAI approaches that is inspired by the sociotechnical nature of these systems. %
To this end, we reviewed state-of-the-art algorithmic and user-centred evaluation strategies as well as frameworks built on top of them. %
This analysis allowed us to identify multiple unaddressed deficiencies in this area: inconsistent evaluation findings, neglected explainability contexts, oversimplified view of the human explanatory process, misconstrued structure of XAI systems as monolithic entities and overlooked significance of the varied quality of the explained models. %
Based on these observations, we proposed to: %
\begin{enumerate*}[label=\protect\cir{\arabic*}]
\item
identify functionally-independent components of explainability systems and map their dependencies; %
\item
evaluate these building blocks on multiple levels -- individually, in modules and holistically; and %
\item
explicitly account for the anticipated operational context during validation. %
\end{enumerate*}

Within this framework, (numerical) \emph{explanatory insights} output by XAI tools can be assessed algorithmically with respect to their technical and user-centred desiderata, whereas \emph{explanatory artefacts} and their \emph{communication processes} can be appraised with user-based evaluation protocols in a well-defined operational context. %
Such an approach promises to streamline the evaluation process and give rise to a comprehensive repository of thoroughly-tested XAI building blocks that can be used to compose bespoke explainers that are trustworthy and suitable for the task at hand. %
While our framework does not provide a prescriptive validation workflow, it offers a principled procedure to reason about XAI evaluation strategies and encourages best practice. %
In future work, we envisage expanding, formalising and refining our approach by proposing relevant templates, protocols and user study designs. %

\renewcommand{\acksname}{Acknowledgements}
\begin{acks}
This research was supported by the Hasler Foundation, grant number 23082. %
\end{acks}

\bibliographystyle{ACM-Reference-Format}
\bibliography{chiea24-571}

\end{document}